\documentclass[%
 aip,
 amsmath,amssymb,
 reprint,%
]{revtex4-1}

\usepackage{graphicx}
\usepackage{dcolumn}
\usepackage{bm}

\usepackage[utf8]{inputenc}
\usepackage[T1]{fontenc}
\usepackage{mathptmx}
\usepackage{etoolbox}

\usepackage{xcolor}
\usepackage[normalem]{ulem}

\makeatletter
\def\@email#1#2{%
 \endgroup
 \patchcmd{\titleblock@produce}
  {\frontmatter@RRAPformat}
  {\frontmatter@RRAPformat{\produce@RRAP{*#1\href{mailto:#2}{#2}}}\frontmatter@RRAPformat}
  {}{}
}
\makeatother

\begin{document}

\title[TDS: Semiconductor Characterization]{Accessibility of doping ranges of semiconductors by terahertz spectroscopy}

\author{J. Hennig}
\email{joshua.hennig@itwm.fraunhofer.de}

\affiliation{Fraunhofer Institute for Industrial Mathematics ITWM, Department Materials Characterization and Testing, 67663\,Kaiserslautern, Germany}
\affiliation{Department of Physics and Research Center OPTIMAS, RPTU University Kaiserslautern-Landau, 67663\,Kaiserslautern, Germany}

\author{J. Klier}
\affiliation{Fraunhofer Institute for Industrial Mathematics ITWM, Department Materials Characterization and Testing, 67663\,Kaiserslautern, Germany}

\author{S. Duran}
\affiliation{Fraunhofer Institute for Industrial Mathematics ITWM, Department Materials Characterization and Testing, 67663\,Kaiserslautern, Germany}

\author{M. Kutas}
\affiliation{Fraunhofer Institute for Industrial Mathematics ITWM, Department Materials Characterization and Testing, 67663\,Kaiserslautern, Germany}
\affiliation{Department of Physics and Research Center OPTIMAS, RPTU University Kaiserslautern-Landau, 67663\,Kaiserslautern, Germany}

\author{J. Jonuscheit}
\affiliation{Fraunhofer Institute for Industrial Mathematics ITWM, Department Materials Characterization and Testing, 67663\,Kaiserslautern, Germany}

\author{G. von Freymann}
\affiliation{Fraunhofer Institute for Industrial Mathematics ITWM, Department Materials Characterization and Testing, 67663\,Kaiserslautern, Germany}
\affiliation{Department of Physics and Research Center OPTIMAS, RPTU University Kaiserslautern-Landau, 67663\,Kaiserslautern, Germany}

\author{D. Molter}
\affiliation{Fraunhofer Institute for Industrial Mathematics ITWM, Department Materials Characterization and Testing, 67663\,Kaiserslautern, Germany}

\date{\today}

\begin{abstract}
While established semiconductor measurement techniques such as four-point probe or capacitance-voltage measurements require a physical contact to the material, terahertz spectroscopy is completely contact-free. Its capability to measure the doping of semiconductors is well known, yet the exact doping ranges that are accessible to terahertz spectroscopy are not obvious. Therefore, we introduce a sensitivity metric to clarify whether a semiconductor sample can be characterized in principle by reflection terahertz time-domain spectroscopy. This quantity takes into account the semiconductor material with a certain layer thickness, doping type, and doping level and is based on numerical simulations. In this work, we calculate this sensitivity value for relevant semiconductor materials (SiC, Si, GaN) in realistic layer stacks with up to three layers. It is used to create meaningful heat maps depending on the thicknesses and charge carrier densities of the sample structures of interest. The plausibility of the sensitivity is validated by mapping a variety of measurements with terahertz techniques from us and from other groups onto these heat maps. Based on these, the accessible range of charge carrier densities for terahertz spectroscopy spans roughly from $10^{15}$~cm$^{-3}$ to $10^{20}$~cm$^{-3}$, but with dependencies on material, doping type, and sample thickness. Furthermore, the sensitivity value allows for a substantiated assessment of the possible benefits future improvements of photoconductive antennas and terahertz systems could have, which is demonstrated by simulations based on varied bandwidths.
\end{abstract}

\maketitle



Semiconductors are fundamental materials of the digital age enabling technological advancements in numerous fields ranging from everyday communication over the electrification of transportation, the buildup of renewable energy production, the rapidly increasing use of artificial intelligence to fundamental research. With the growing demand for semiconductor devices going along with these changes, the need for fast and non-destructive characterization techniques of these materials is growing quickly as well. 

Terahertz time-domain spectroscopy (THz-TDS) has proven to be capable of measuring the desired optical and electrical properties of various semiconductors. \cite{van1990carrier, jeon1997nature, jeon1998characterization, jeon1998observation, nashima2001measurement, hangyo2002spectroscopy, herrmann2002terahertz, zhang2003terahertz, nagai2006characterization, guo2009terahertz, hamano2012rapid, hamano2014high, alberding2017direct, kim2017terahertz, hennig2024wide, hennig2025simultaneous} However, it is not readily apparent which materials' doping ranges and layer thicknesses are addressable with THz-TDS. Therefore, we developed a simulation-based tool to identify the accessible doping ranges. Based on simulations of terahertz pulses and the Drude model\cite{yu2005fundamentals, dresselhaus2022solid}, as well as the interaction between a terahertz pulse and a freely designable semiconductor layer stack, it is possible to predict whether THz-TDS is capable of characterizing an arbitrary sample of interest. These simulations provide a better, more descriptive visual understanding of suitable samples for THz-TDS characterization, including limits depending on material properties. 
This has the potential to further establish THz-TDS as a tool in semiconductor science and industry.

\begin{figure*}[htb]
\includegraphics{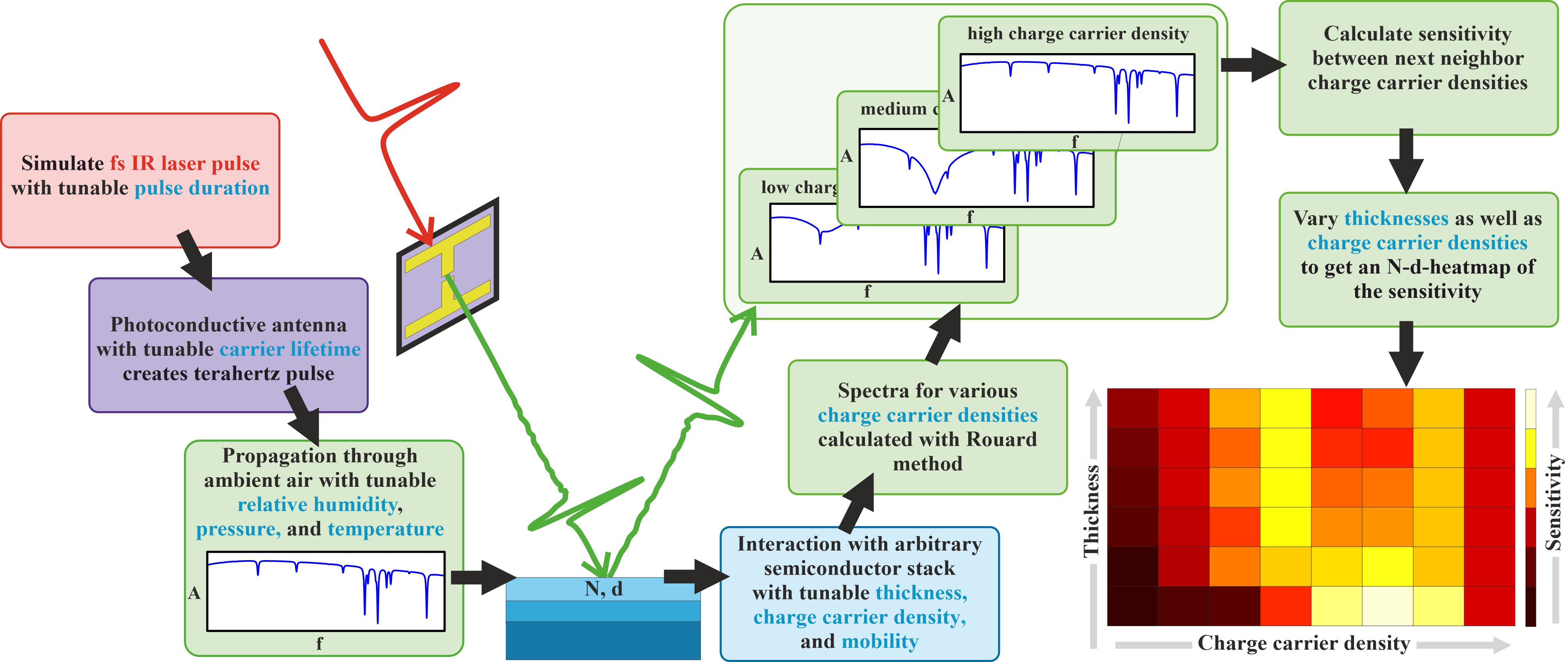}
\caption{\label{fig:1_Schematic_Simulation}Schematic of the simulation steps to calculate the sensitivity. First, an IR fs-laser pulse (red) is calculated, triggering a THz-pulse to be emitted from a photoconductive antenna (purple). After calculation of the propagation in the frequency domain (green), the material interaction is simulated (blue); here, $N$ of one layer of the sample is varied systematically leading to a series of spectra (green). Between the spectra of neighboring $N$ the sensitivity is calculated. The same is repeated for various thicknesses leading to a $d$-$N$-heat map with the colorbar representing the sensitivity of terahertz measurements.}
\end{figure*}

To model semiconductor samples, the Drude model\cite{yu2005fundamentals, dresselhaus2022solid} is used in the way described in supplementary material I. The material parameters used throughout this work are listed in supplementary material II. This approach has already been used successfully to evaluate THz-TDS measurements of semiconductors to retrieve the resistivity \cite{hennig2024wide} or charge carrier density \cite{hennig2025simultaneous} $N$ of the samples under test. In this work, we turn the tables and apply this evaluation to numerically simulated THz-TDS measurements of samples with various thicknesses and charge carrier densities. A comparison of this multitude of resulting spectra allows to narrow down the ranges of these properties, which one can expect a THz-TDS measurement to be clearly distinguishable in. This distinctiveness is expressed by a single sensitivity value. Hence, it helps to clarify whether a doping range is accessible by THz-TDS. 

\begin{figure*}
\includegraphics{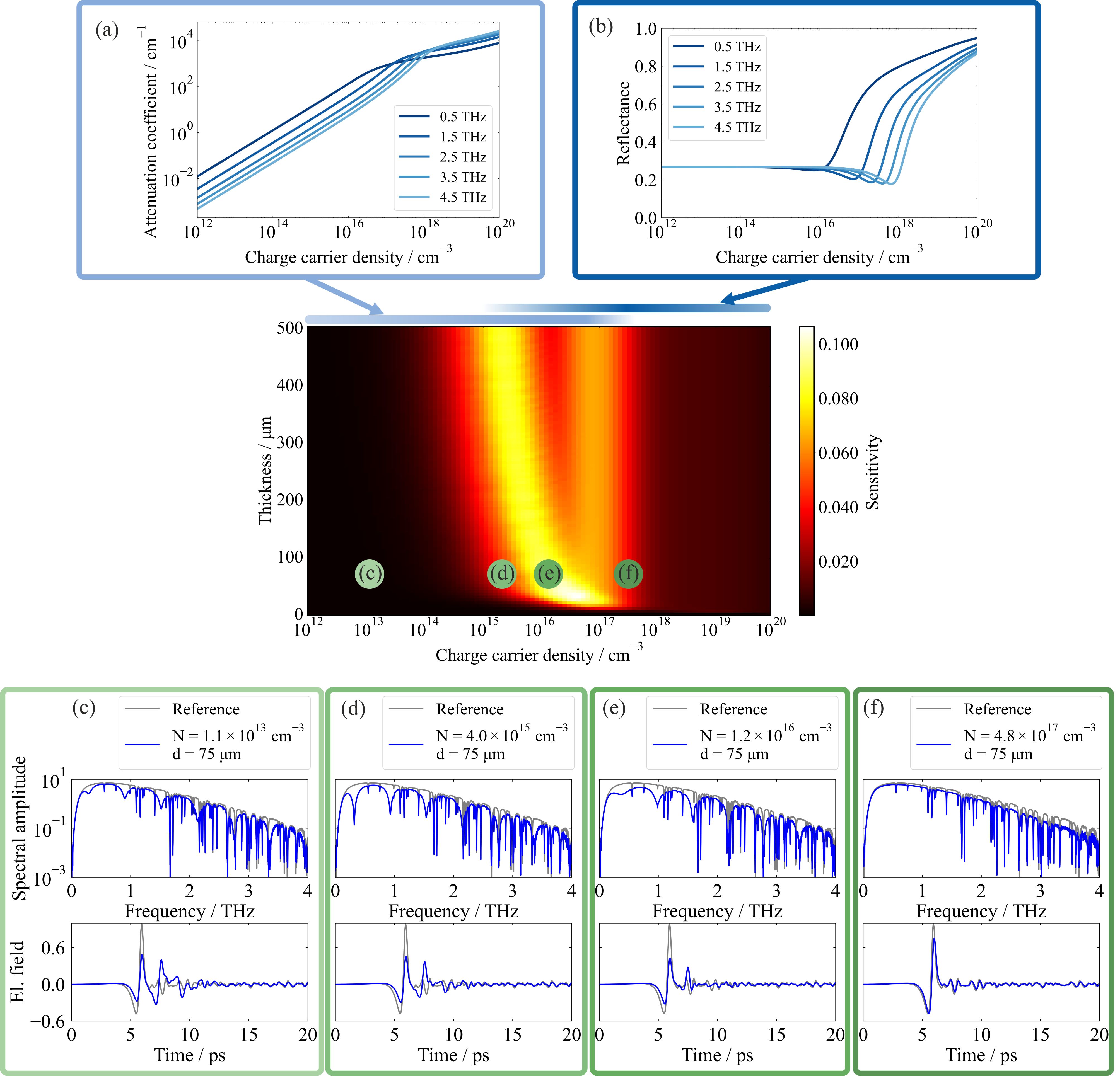}
\caption{\label{fig:2_Explanation_Heatmap} Sensitivity heat map (center) for 3-layer SiC samples simulated with the methods described in Fig. \ref{fig:1_Schematic_Simulation}. Further explanatory graphs show the attenuation coefficient (a) and the reflectance (b) for SiC as a function of the charge carrier density for the relevant terahertz frequencies. The ranges of $N$, which they each influence the sensitivity dominantly in, are marked above the heat map. In (c) - (f) corresponding spectra and waveforms in blue for the respectively marked $d$- and $N$-values in the heat map are plotted together with corresponding references in grey, exemplarily showing the impact of an increasing attenuation and reflectance.}
\end{figure*}

The complete path of simulations and numerical calculations is depicted schematically in Fig. \ref{fig:1_Schematic_Simulation}. A numerical simulation of the infrared fs-laser pulse (red-colored) with tunable pulse duration used in THz-TDS systems is carried out to begin the simulation. For a standard THz-TDS system, we assume a pulse duration of 100~fs. Next, the photoconductive antenna (purple-colored) is simulated with a tunable carrier lifetime. Here, it is assumed to be 0.35~ps.\cite{deumer2025advancing, han2025advancements} Then the time-domain pulse is transferred to the frequency domain (green-colored) by a fast Fourier transform. A high frequency resolution of 1 GHz is used, realized by choosing the time axis in the time-domain sufficiently long, which can realistically be achieved in a real-world measurement by zero filling. Next, a high pass filter is applied to model losses of the lower frequencies due to the aperture of the optics guiding the terahertz radiation. Further, air propagation is modeled based on data of water vapor absorption lines from the HITRAN database. \cite{rothman2013hitran2012} The complex index of refraction is calculated with the Kramers-Kronig relations. \cite{herrmann2012combination} Here, we assume the following tunable values: 0.5~m propagation length, 30~\% relative humidity, 1~bar air pressure and a temperature of 22~°C. The idea behind taking the impact of water vapor absorption into account as an option in these simulations is that not all measurements can be operated in a purged environment; this could be specifically relevant for possible applications in industrial environments. Another photoconductive antenna with a slightly shorter carrier lifetime of 0.15~ps, which is essential to allow sampling \cite{globisch2017iron}, is simulated to model the detection. More details to this approach to simulate the spectrum of a terahertz pulse are described elsewhere. \cite{molter2017interferometry} Based on these steps, a reference pulse is simulated. Adding sufficient white noise to this spectrum leads to a realistic bandwidth of about 4~THz and a signal-to-noise ratio of about 60~dB. A comparison of a pulse and the corresponding spectrum simulated by this procedure with one actually measured can be found in the supplementary material IV.

With such a pulse, the electromagnetic wave-material interaction (blue-colored) is simulated. Here, the Drude model-based description is used in combination with the Rouard method\cite{krimi2016non} to calculate the reflected terahertz radiation.
A stack of arbitrary semiconductor materials can be simulated with n-type or p-type doping with an arbitrary number of layers with tunable thicknesses, charge carrier densities, and corresponding mobilities. In a first step, $N$ of the layer of interest is varied resulting in a number of spectra (green-colored silhouette) corresponding to samples with respective carrier densities. Between each two neighboring complex spectra, consisting of $l$ amplitude values, $a_i$ and $b_i$, representing a single frequency each, the sensitivity is calculated using a root-mean-square criterion
\begin{equation}
    \text{Sensitivity} \equiv  \sqrt{\frac{\sum_{i=1}^{l}(|a_i|-|b_i|)^2}{l}}.
    \label{eq:Sensitivity}
\end{equation}
This sensitivity in equation (\ref{eq:Sensitivity}) serves as a measure for the dissimilarity of the two respective spectra as a function of the charge carrier density for a sample with a given thickness of the layer examined. Next, the thickness of this layer is varied, and the sensitivity for each $N$ is determined for each thickness individually. The results are plotted on a descriptive heat map with the thickness and charge carrier density on the two axes and the sensitivity as colorbar. A higher sensitivity value indicates that a THz-TDS measurement is more likely to allow the determination of $N$, as the spectrum of a measurement of a given sample is more distinct compared to slightly different samples.

\begin{figure*}
\includegraphics{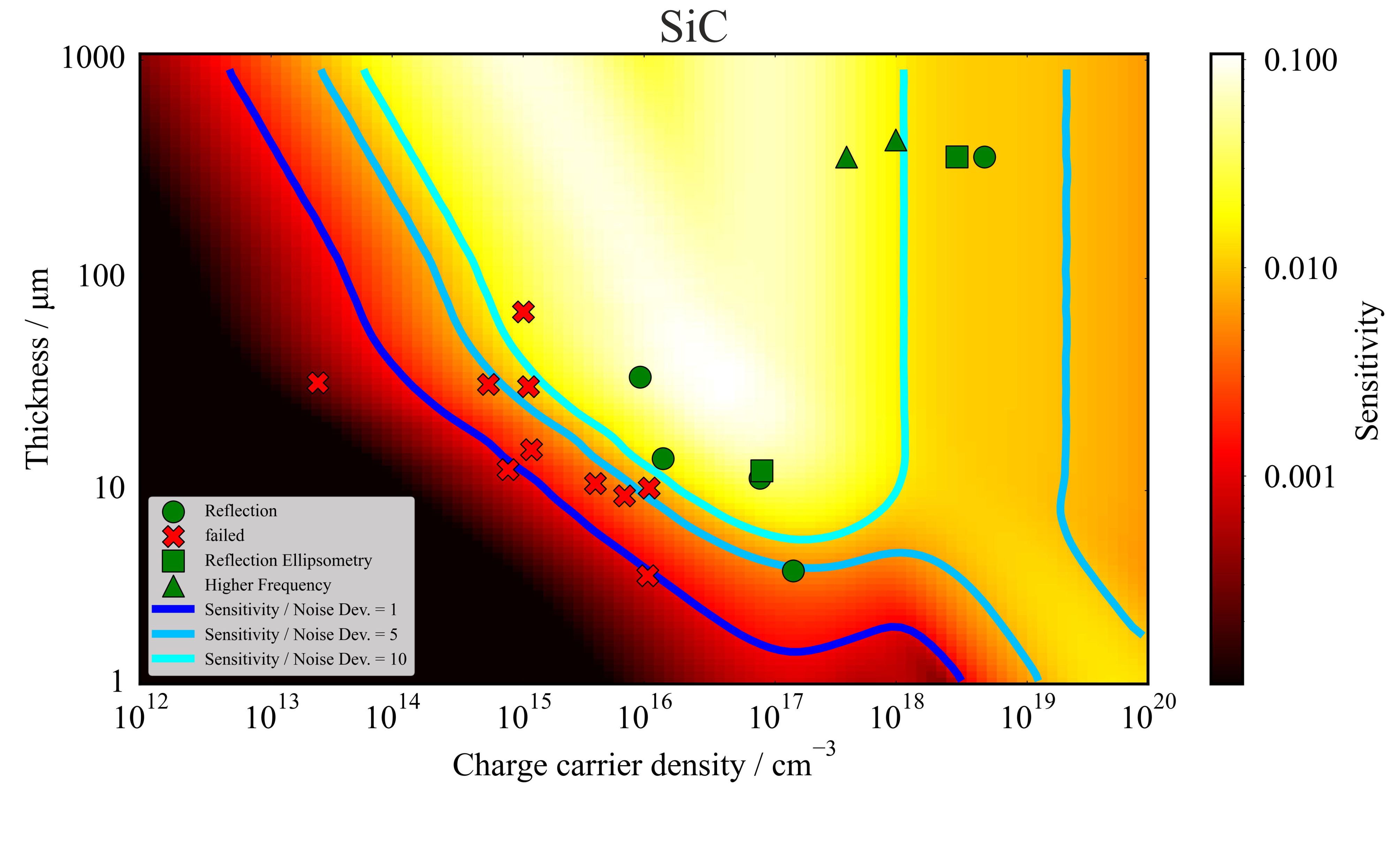}
\caption{\label{fig:3_SiC_Heatmap}Sensitivity heat map for 3-layer n-type SiC samples with logarithmic $d$-axis and colorbar. Further, the lines of noise devided by the sensitivity value equaling 1, 5, and 10 are sketched and samples measured in peer-reviewed papers as well as ones, which a model-based characterization failed for, are plotted.}
\end{figure*}

The resulting sensitivity heat map for a three-layer sample of 4H-SiC is shown in the middle of Fig. \ref{fig:2_Explanation_Heatmap} for 100 logarithmically scaled $N$-values and 100 linearly scaled $d$-values. The n-type sample simulated here consists of a highly-doped substrate layer (350~µm, $2.5\times10^{18}$~cm$^{-3}$), a buffer layer (1~µm, $1\times10^{18}$~cm$^{-3}$), and a tunable epi-/top-layer with thicknesses in the range of 1~µm to 500~µm and charge carrier densities in the range of $1\times10^{12}$~cm$^{-3}$ to $1\times10^{20}$~cm$^{-3}$. A typical epilayer sample, that could be found in an intermediate process step in fabrication of SiC power devices, would be located in the range of thin layer thicknesses and medium carrier densities.\cite{kimoto2022high} In the heat map, two main branches appear with high values of sensitivity. The one towards lower charge carrier densities is mainly due to an increase in the attenuation coefficient, plotted in (a) for the $N$-values and the relevant terahertz frequencies. It is dominating for values of $N$ in the range from about $1\times10^{14}$~cm$^{-3}$  to $1\times10^{17}$~cm$^{-3}$. This branch stems from losses due to absorption and scattering during propagation in the material. Therefore, it undergoes a shift towards lower $N$-values with increasing thickness. For even lower values of $N$~<~$1\times10^{14}$~cm$^{-3}$, the samples appear transparent due to small attenuation. The other branch is fixed around $N$~$\approx$~$1\times10^{17}$~cm$^{-3}$. Around this value, the reflectivity (b) for the terahertz frequencies is at the Drude edge, dropping first before increasing strongly. \cite{yu2005fundamentals} According to equation S2 in the supplementary material, the plasma frequency increases with an increase in doping concentration due to more free charge carriers contributing to a higher Coulomb force restoring the charge carriers against their displacement. In this model, the assumed harmonic oscillator frequency therefore increases crossing the terahertz frequency range. This strongly affects the measured signal with the material becoming opaque. The changes in reflectivity dominate for higher values of $N$. However, the further above $N$~$\approx$~$1\times10^{17}$~cm$^{-3}$ the samples are, the smaller the changes in the signal become leading to a monotonous decline in sensitivity.

In Fig. \ref{fig:2_Explanation_Heatmap} (c) - (f), the respective spectra and time-domain waveforms for four $d$-$N$-combinations in the heat map  are plotted exemplarily to illustrate these effects. In the background of each of these plots, the corresponding simulated reference is plotted for comparison. For the least doped sample in (c), the attenuation losses are small. The spectrum shows Fabry-Pérot dips and the time-domain amplitude is split up between the front-side reflection and multi-reflections inside the sample. With increasing $N$-values the attenuation increases and the Fabry-Pérot dips become deeper because the sum of the amplitudes of the multi-reflection peaks in the time-domain better matches the amplitude of the front-side reflection. While the latter undergoes only a phase shift of $\pi$ from the reflection at the optically thicker material, the others undergo phase shifts of multiples of $2\pi$ because of propagation and reflection at their frequencies. Therefore, they interfere destructively with the front-side reflection. This is shown in (d). Further increasing the doping, and hence the attenuation, leads to the epilayer becoming less transparent, making the multi-reflections vanish (e), which corresponds to the Fabry-Pérot dips vanishing, too. Once they disappear completely (f), a spectrum without Fabry-Pérot resonances is left corresponding to the remaining front-side reflection. Increasing $N$ further only leads to slight changes in the amplitude. So, the resulting heat map is affected by the depth and shape of the Fabry-Pérot resonances as well as the total changes in amplitude across the whole spectrum and the dispersion of the underlying attenuation and reflection at each interface. While the total spectral amplitude may seem similar in Figs. 2 (c) – (f) due to the highly reflective substrate below the buffer- and epilayer, they differ relevantly due to slight changes in the material parameters. The rate of this local alteration is described by the sensitivity value shown in the heat map.

A possible negative impact of the water vapor absorption lines in general is manageable in this consideration. Due to the high frequency resolution, the absorption lines appear smooth. Obviously they stay at the same frequency position for each simulation making them cancel out each other and thereby their impact for the most part in the comparison of two spectra. Furthermore, the amplitude lost to them is relatively small compared to the amplitude in a whole spectrum.

While the two distinct branches of the heat map in Fig. \ref{fig:2_Explanation_Heatmap} help to physically explain the effects on the measured signal, this depiction does not reveal where the boundaries of a THz-TDS measurement are, yet. Therefore, an equivalent simulation is plotted as a heat map in Fig. \ref{fig:3_SiC_Heatmap}, but with a logarithmically scaled colorbar to allow for a better distinction between the lower values of the sensitivity. Additionally, the thickness is logarithmically scaled, giving better insight into the range below 100~\textmu m, 
and extended to 1000~\textmu m. 

\begin{figure*}
\includegraphics{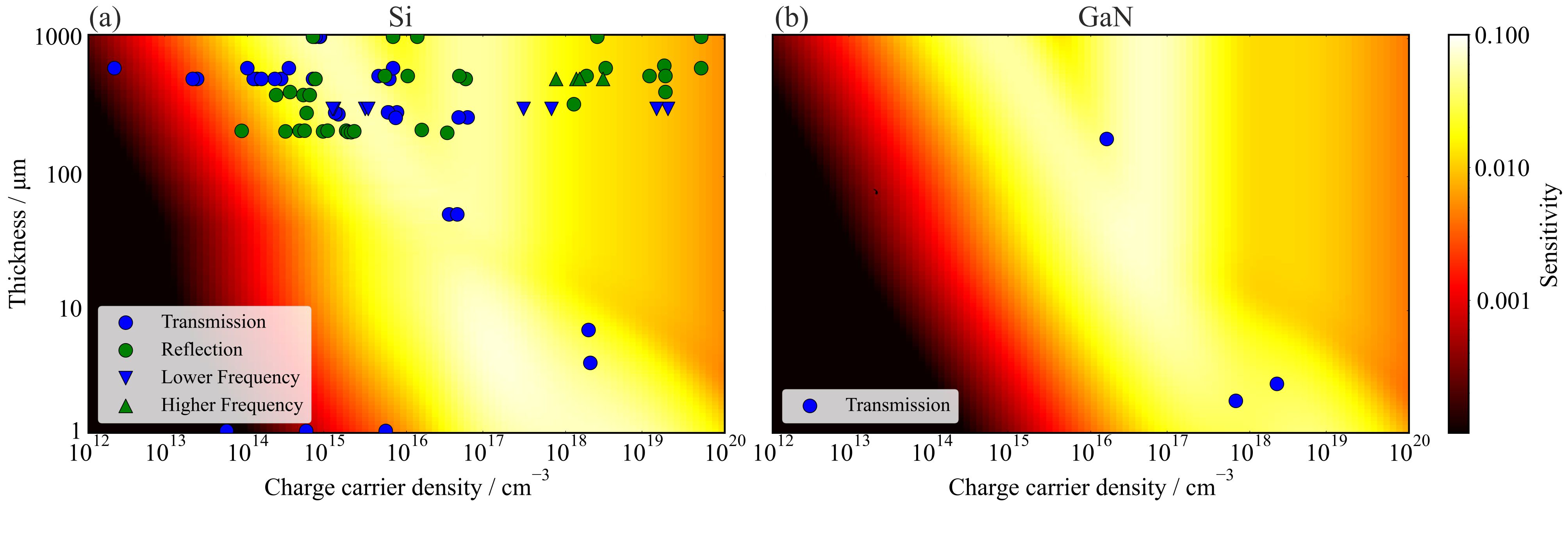}
\caption{\label{fig:4_Si_GaN_Heatmap} Sensitivity heat maps for (a) 1-layer n-type Si and (b) 2-layer n-type GaN on top of sapphire, both with samples plotted, which were measured in peer-reviewed publications with terahertz spectroscopy techniques.}
\end{figure*}

The contour lines in tones of blue are based on additional simulations of the heat map with random white noise. At each $d$-$N$-position, the spectra are simulated with individual random noise of the same level. Then, the noise standard deviation is calculated. Finally, the sensitivity at each position is divided by the respective noise deviation. The lines indicate the heat map positions, at which this ratio equals 1, 5, or 10. So, these iso-sensitivity-noise-lines indicate the $d$-$N$-combinations, at which the change in the signal due to the sample properties is either equal to the effect of the noise or 5 or 10 times higher. This shows that a doping range from about $1\times10^{15}$~cm$^{-3}$ to $1\times10^{20}$~cm$^{-3}$ is accessible. The exact numbers depend on the noise level and on the layer thickness. For very high $N$-values above $1\times10^{19}$~cm$^{-3}$ and thicknesses below 10~\textmu m, an increase in sensitivity falls out of the line. This case is explained by the samples in that range being so thin that they allow for a significant percentage of the terahertz radiation to be transmitted through the epilayer despite the high attenuation. This corresponds to the extension of the attenuation-related branch in the heat map. Therefore, the epilayer is probed more precisely than without the reflection signal from the next interface.

To put these theoretical calculations into perspective to measurements of various terahertz spectroscopy approaches reported in literature, the green dots represent samples measured with THz-TDS in reflection geometry \cite{hennig2025simultaneous}, the green squares represent samples measured with terahertz reflection ellipsometry \cite{nagashima2013characterization, agulto2025wafer}, the green triangles pointing upward represent samples measured in reflection with higher terahertz frequencies \cite{hamano2014high}, and the red crosses represent samples that we were unable to successfully characterize in our own studies so far. \cite{hennig2025simultaneous}

Although not all these samples may agree perfectly with the 3-layer stack simulated, they are of the same material, with variations in the doping of their topmost layer. Therefore, they are suitable for this analysis. From the distribution of the successful and unsuccessful characterizations of the samples, one can say that a sensitivity value in the same order as the noise deviation typically is not good enough to allow a characterization. One needs at least a 5- or better 10 times higher sensitivity value to be confident in an electrical characterization. Also, from our studies we came to the conclusion that a good knowledge of the layer thicknesses of the sample is essential. \cite{hennig2025simultaneous} A lack of this knowledge might explain some of our failed measurements because the buffer layer thickness is not convincingly characterized -- especially for the thickest sample corresponding to a red cross, we suspect the buffer layer to be thicker than assumed originally.

A similar simulation and literature study for silicon and gallium nitride is summarized in the two heat maps in Fig. \ref{fig:4_Si_GaN_Heatmap}. For Fig. \ref{fig:4_Si_GaN_Heatmap} (a) a single n-doped silicon layer is simulated. Analogously, for Fig. \ref{fig:4_Si_GaN_Heatmap} (b) a two-layer system of a top-layer of n-doped GaN on a sapphire substrate is simulated. Again, in both graphs, the same two branches of high sensitivity as in Fig. \ref{fig:1_Schematic_Simulation} are dominating, associated with attenuation inside the material and a rise in reflectivity of the topmost semiconductor layer. For the silicon samples, green dots correspond to THz-TDS measurements of silicon samples taken from literature. \cite{jeon1998characterization, nashima2001measurement, nagai2006characterization, hennig2024wide} The green triangle pointing upwards represents samples investigated with higher terahertz frequencies in reflection. \cite{hamano2012rapid} Blue dots stand for THz-TDS transmission measurements. \cite{van1990carrier, jeon1997nature, jeon1998observation, hangyo2002spectroscopy, herrmann2002terahertz,  nagai2006characterization, alberding2017direct} Finally, blue triangles pointing downwards stand for samples measured with lower terahertz frequencies in transmission geometry. \cite{kim2017terahertz} Because single-layer Si wafers are investigated in many of these works, the simulations, which the sensitivity is based on, are also for single-layer Si. The differences of n-doped and p-doped samples in the effective masses and mobilities only cause slight variations in the sensitivity heat map -- therefore, this distinction is neglected for a better overview covering this material completely in one graph. The transmission measurements, however, are obviously not in reflection geometry, yet they are based on the same physical laws and included for completeness. It is to be expected that most of them lie in the range of lower doping < $1\times10^{17}$~cm$^{-3}$ to have a relevant percentage of terahertz radiation transmitted through the sample. For thin layers (< 7~\textmu m) \cite{alberding2017direct}, the extensions of the branch of dominating attenuation in the higher doped ranges is explainable due to thin samples. Interesting are the low-frequency measurements of thicker and higher doped samples. \cite{kim2017terahertz} However, they report only very little signal, and lack a complete model-based evaluation. The same can be said about the very thin (< 1~\textmu m) doped silicon layers that were produced with implantation. \cite{herrmann2002terahertz} Beside these special cases, all these measurements from literature are in good agreement with the assessment that a high sensitivity allows the samples to be characterized by terahertz spectroscopy.

An analogous analysis is done for Fig. \ref{fig:4_Si_GaN_Heatmap} (b). GaN has not been studied as extensively, yet. Therefore, there are less samples of doped GaN investigated by terahertz spectroscopy. To the best of our knowledge, those reported in literature are measured in transmission \cite{guo2009terahertz, zhang2003terahertz}, and are mapped onto the sensitivity heat map. It is evident that these characterizations are well inside the ranges, which the terahertz measurements are sensitive for. Given that these are samples on a sapphire substrate \cite{guo2009terahertz} or at least grown on one \cite{zhang2003terahertz}, the heat map is simulated for the corresponding two-layer system with the substrate material differing from the heteroepitaxially grown epilayer. This shows the versatility of this simulation method to model almost arbitrary semiconductor material stacks. Further, the samples in the range with high doping concentration > $1\times10^{18}$~cm$^{-3}$ and thicknesses below 10~\textmu m could again be measured in transmission geometry, verifying the explanation that reflection measurements in this range probe the sample to the next interface.

All the above-mentioned simulations are based on the assumption of a terahertz pulse with a bandwidth of about 4~THz. An analysis of the impact of different bandwidths provided by more improved photoconductive antennas \cite{dohms2024fiber} can be found in the supplementary material III.

In conclusion, we carried out extensive simulations of reflection terahertz time-domain spectroscopy of semiconductor samples covering wide ranges of thicknesses from 1~\textmu m to 1000~\textmu m and charge carrier densities from $1\times10^{12}$~cm$^{-3}$ to $1\times10^{20}$~cm$^{-3}$. Exemplarily, the materials Si, GaN, and SiC in respective stacks from 1 to 3 layers were considered. The sensitivity value as a measure of the characterizability of a certain sample was introduced, indicating how much two spectra differ from one another. Using this quantity, descriptive heat maps visualize the combinations of thicknesses and charge carrier densities accessible by terahertz spectroscopy. 

Roughly, $N$-ranges from about \mbox{$1\times10^{15}$~cm$^{-3}$} to \mbox{$1\times10^{20}$~cm$^{-3}$} are accessible, but with a strong dependence on the layer thickness. The material and doping type only play minor roles in setting these boundaries. Generally, one can say that thicker samples can have a smaller charge carrier density to still be accessible. A special feature is that for even higher doping levels, a thinner sample could be advantageous because it might still allow to be penetrated completely. This assessment is validated by a literature analysis of measurements with various terahertz spectroscopy techniques of the investigated semiconductors. Finally, the flexibility of this simulation approach is utilized to vary the bandwidth of the terahertz spectra and thereby underline that the ranges found are fundamentally based on the frequency-dependent material properties.

As an outlook for possible future work, the presented simulation approach could be extended to take transmission measurements into account as well as to consider ferromagnetic materials with a permeability $\tilde{\mu} \neq 1$, as it has been shown before that such materials can be characterized by a combination of transmission and reflection TDS measurements.\cite{papari2025accurate} This would, however, require implementing an additional evaluation method for the transmission measurements, e.g. a transfer matrix method. Further, the combination of a transmission and a reflection measurement would have to be implemented numerically as well and combined with an extended formula for a sensitivity value taking both of them into account. While this could be an interesting analysis, it is outside the scope of this work.

\section*{Supplementary material}
See the supplementary material for more details and explanations concerning the following topics: Section I of the supplementary material covers the Drude model and which formulas were used exactly in our numerical calculations. Section II lists the material parameters taken from literature and assumed for the respective semiconductor materials SiC, Si, and GaN. Section III contains an analysis of the impact of different realistic bandwidths of photoconductive antennas on the simulation results. Section IV shows a comparison of a simulated pulse and corresponding spectra to one actually measured. 

\begin{acknowledgments}
This project is supported by the Federal Ministry for Economic Affairs and Climate Action (BMWK) on the basis of a decision by the German Bundestag.
\end{acknowledgments}

\section*{Data Availability Statement}

The data that support the findings of this study are available from the corresponding author upon reasonable request.

\section*{Disclosure}

The authors have no conflicts to disclose.

\clearpage 
\section*{Supplementary material}

\section{Drude model}

Electromagnetic waves can couple with the available charge carriers in semiconductors, electrons or holes, whose numbers are typically controlled by the respective doping concentration. Additionally, the optical phonons of most semiconductors have frequencies higher than the frequency range of 0.1 to 4 THz, which is typically used in THz-TDS measurements. Therefore, the simple Drude model \cite{yu2005fundamentals, dresselhaus2022solid} can be used to describe conduction in this case, neglecting any Lorentz terms from phonon resonances. According to the Drude model, the dielectric function of a semiconductor $\tilde{\epsilon}$ is given by
\begin{equation}
    \tilde{\epsilon} = \epsilon_{\text{BG}}-\frac{\omega_{\text{p}}^{2}}{\omega(\omega+i\Gamma)}
    \label{eq:dielectric_function_Drude}
\end{equation}
with the background dielectric constant $\epsilon_{\text{BG}}$, the angular frequency $\omega$, the imaginary unit $i$ and the damping frequency $\Gamma$. $\omega_{\text{p}}$ is the plasma frequency, which can be expressed by 
\begin{equation}
    \omega_{\text{p}}^{2} = \frac{Nq^{2}}{\epsilon_{0}m_{\text{e}}m^{*}}
    \label{eq:plasma_frequency}
\end{equation}
with the charge carrier density $N$, the elementary charge $q$, the vacuum permittivity $\epsilon_{0}$, the electron mass $m_{\text{e}}$, and the effective mass for conductivity $m^{*}$.  Inserting equation (\ref{eq:plasma_frequency}) into equation (\ref{eq:dielectric_function_Drude}) and replacing the damping with an expression for the mean free time $\tau = \frac{m^{*}m_{\text{e}}\mu}{q}$, one receives the formula 
\begin{equation}
    \tilde{\epsilon}(N, \omega) = \epsilon_{\text{BG}}-\frac{Nq^{2}}{\epsilon_{0}m_{\text{e}}m^{*}\omega \big (\omega+i\frac{q}{m_{\text{e}}m^{*}\mu(N)} \big )}
    \label{eq:dielectric_function_Drude_N_dependend}
\end{equation}
for the dielectric function. \cite{hennig2024wide, hennig2025simultaneous} Here, the charge carrier mobility $\mu$ is expressed depending on the carrier density $N$ as it can be modeled for many semiconductors by a Caughey-Thomas model \cite{caughey1967carrier}. Equation (\ref{eq:dielectric_function_Drude_N_dependend}) is used throughout this work to describe the respective semiconductor samples. The advantage of this expression is that material-specific quantities $\epsilon_{\text{BG}}$, $m_{\text{e}}$, and $\mu(N)$ can be taken from literature as well as the constants q and $\epsilon_{0}$. So, $\tilde{\epsilon}$ only depends on $N$ -- besides the frequency-dependence allowing dispersion to be accounted for.

The interaction between the terahertz radiation and the material is described by the Fresnel equations. Further, the Rouard method is used to describe the transfer function of semiconductor stacks with multiple layers for the terahertz simulations. This evaluation method as well as the model are described in more detail in.\cite{hennig2025simultaneous}

In certain instances, the Drude-Smith model is also used to describe the electrodynamic response of semiconductors. However, it is especially useful in the context of samples, in which scattering plays a major role, e.g. for nanocrystals or polycrystalline materials.\cite{zajac2014thz} In this work, we are considering bulk, single-crystal materials such as semiconductor wafers or epitaxially grown epilayers, that are homogenously structured and exhibit little defects. A discussion of both cases of the Drude and the Drude-Smith model can be found for silicon exemplarily.\cite{titova2016ultrafast}  Therefore, at the length scales of the mean free path of the charge carriers, the materials considered are homogenous and therefore the Drude model can be applied. An adoption of the Drude-Smith model to our simulation approach is in general be possible. However, this would inevitably require changes to the model simulated and add at least one additional variable to the evaluation with the c parameter of the Drude-Smith model. This variable could be considered unknown for a respective sample and would therefore require either further calibration or a reliable value from literature.

\section{Material parameters}

The material parameters listed in table \ref{tab:table1} are used for the simulations of the respective materials throughout this work. For SiC the polytype 4H-SiC is assumed, and for GaN the hexagonal structure is assumed. The parameters needed are the background dielectric constant $\epsilon_\text{BG}$, the mobility $\mu$, and the effective mass $m^{*}$.

\begin{table}[h]
\caption{\label{tab:table1}Material parameters used for SiC, Si, and GaN. 
\protect\\ Alt Text: The table contains three rows for the materials SiC, Si, and GaN. The following quantities are listed in the columns: the background dielectric constant, the charge carrier mobility, and the effective mass.}
\begin{ruledtabular}
\renewcommand{\arraystretch}{1.5}
\begin{tabular}{cccc}
  
 & $\epsilon_\text{BG}$ & $\mu$($N$) & $m^{*}$ \\
\hline
SiC\cite{naftaly2016silicon, habib2012finite, burin2024tcad, ishikawa2021electron} 
& 9.91
& $40+\frac{910 - 40}{1 + (N/(2\times10^{17})^{0.76})}$  
& 0.404 \\
\hline
Si\cite{dai2004terahertz,caughey1967carrier,jeon1997nature} 
& 11.68
& $65+\frac{1330 - 65}{1 + (N/(8.5\times10^{16})^{0.72})}$ 
& 0.26 \\
\hline
GaN\cite{meneghini2021gan, arakawa2017electrical}
& 8.9
& $130+\frac{1035 - 130}{1 + (N/(3\times10^{17})^{1.0})}$
& 0.20\\
\end{tabular}
\end{ruledtabular}
\end{table}

\section{Impact of the bandwidth}

\begin{figure}[bp]
\includegraphics{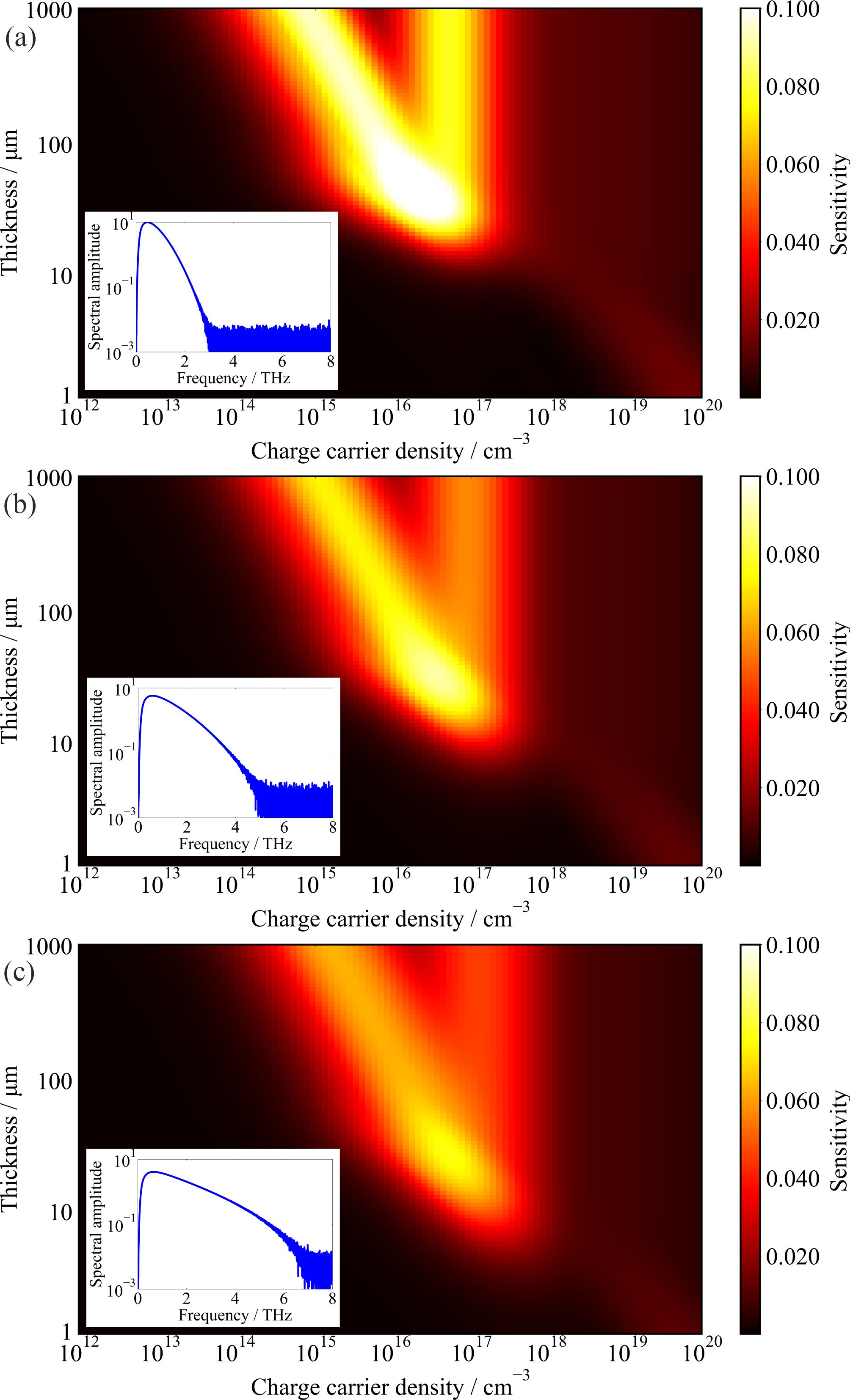}
\caption{\label{fig:5_SiC_varied_pulse_lengths} Sensitivity heat maps based on simulations of terahertz pulses with varied bandwidths, roughly (a) 2~THz, (b) 4~THz, and (c) 7~THz. The insets show the respectively used terahertz spectra of the incoming terahertz pulse used for each simulation.
\protect\\ Alt Text: Simulaed d-N-heat maps of the sensitivity are plotted are different simulated bandwidths of about 2 THz, 4 THz, and 7 THz. Again, two branches can be identified, that are slightly broadened with higher bandwidths. In each heat map an inset shows the underlying simulated reference spectrum.}
\end{figure}

The simulations in this work so far are based on the assumption of a terahertz pulse with a bandwidth of about 4~THz. Fig. \ref{fig:5_SiC_varied_pulse_lengths} shows a comparison of sensitivity heat maps, again for a 3-layer SiC sample as in Fig. 2 in the manuscript. While keeping the total amplitude of the fs-pulse constant, the bandwidths are varied by adjusting the pulse durations in between the three simulations starting from about (a) 2~THz, corresponding to photoconductive antennas used in early pioneering work. \cite{van1990carrier} In Fig. \ref{fig:5_SiC_varied_pulse_lengths} (b) the same bandwidth of 4~THz as used in all other simulations before is used again as a nowadays commonly-used terahertz spectrum for comparison. Finally, in Fig. \ref{fig:5_SiC_varied_pulse_lengths} (c) a  recently reported wide-bandwidth photoconductive antenna with about 7~THz is simulated. \cite{dohms2024fiber} The insets in the heat maps show the respective simulated reference spectra used. Comparing the three heat maps, one can see that a short bandwidth in combination with a higher amplitude at the lower frequencies leads to a rise in the sensitivity in both branches. With an increase in bandwidth, both branches generally remain at the same $d$- and $N$-positions in the heat map. The attenuation-related branch just experiences a slight expansion towards lower $N$-values and the reflection-related branch experiences an expansion towards higher doping levels, which both is in accordance to the frequency shifts of the attenuation and reflectance, see Fig. 2 (a) and (b) in the manuscript. 

This shows that the doping ranges, which terahertz spectroscopy is a suitable measurements technique to characterize semiconductors in, are given by the physical constraints of the radiation used, rather than its spectral width. A moderate and realistic increase in bandwidth does not promise to overcome these constraints; it just helps to slightly increase the accessible ranges, but does not provide access to completely differing orders of magnitudes in doping or thickness. Depending on the exact sample properties one is interested in, it could even be the better choice to use photoconductive antennas with a smaller bandwidth, but a higher peak dynamic range instead. 

\section{Comparison of simulation and measurement}
\begin{figure}[htbp]
\includegraphics{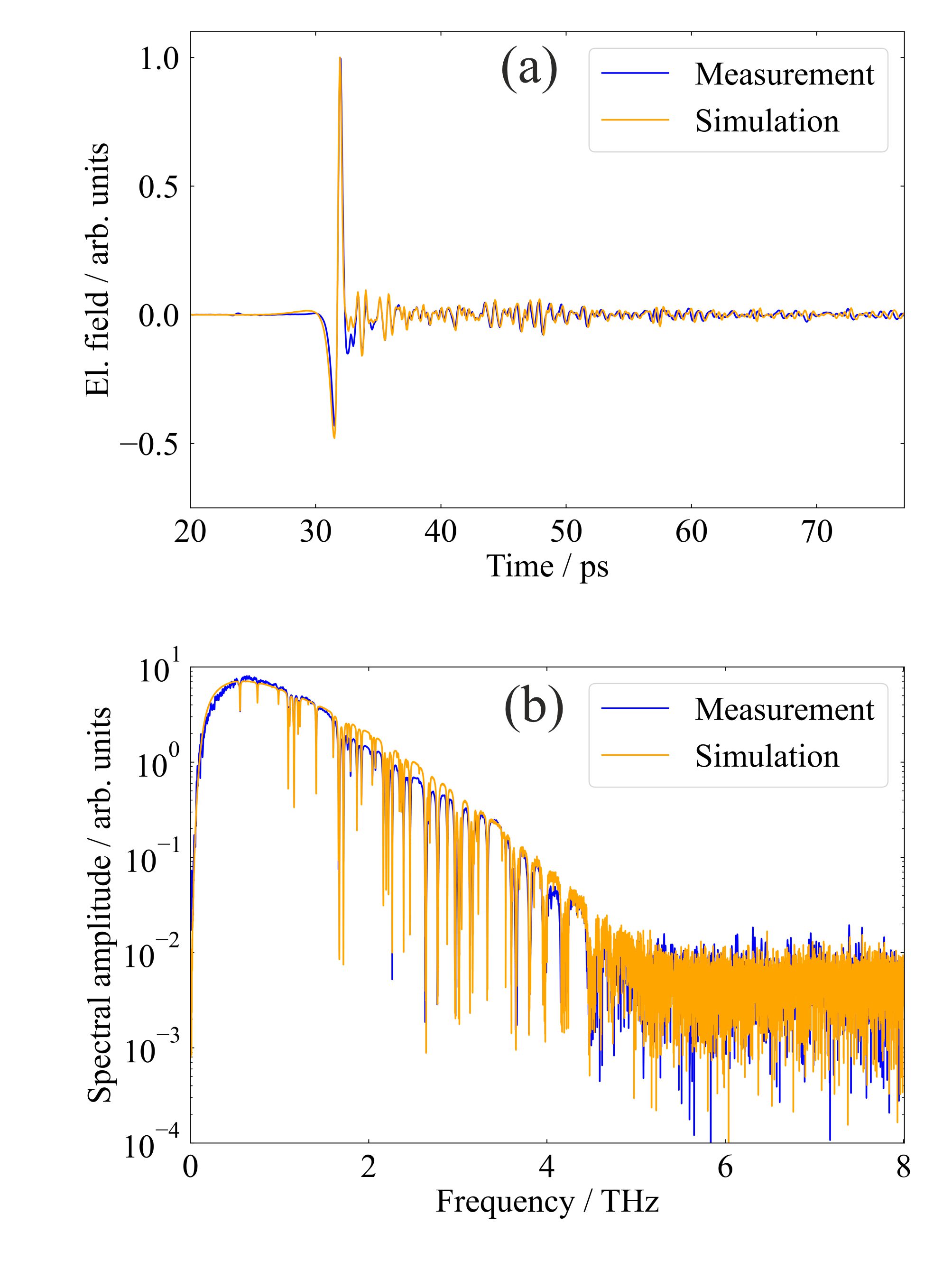}
\caption{\label{fig:6_SiC_varied_pulse_lengths} Comparison of simulations in orange with real measurements in blue (a) of time-domain waveforms and (b) the corresponding spectra.
\protect\\ Alt Text: Two typical ps-short terahertz pulses are shown in (a) with one of them being a real reference measurement of a metal plate and one of them a simulated reference. In (b) the corresponding frequency-domain spectra are shown for these two cases. In both plots, the simulation agrees well with the measurement.}
\end{figure}

In order to prove that the simulations throughout this work offer realistic terahertz pulses and spectra to work with, a comparison with a real measurement is helpful. Therefore, Fig. 2 shows time-domain waveforms (a) of a real reference measurement of a metal plate in blue and of a simulation in orange as used as a reference throughout the simulation procedure. Fig. \ref{fig:6_SiC_varied_pulse_lengths} (b) shows the corresponding frequency-domain amplitude spectra.

The measurement and simulation agree well with each other, especially in the frequency range of highest amplitude below 2 THz. Between about 2 THz and 3 THz, measurement and simulation differ slightly in amplitude before they are in good agreement again above 3 THz before the amplitude falls below the noise level slightly above 4 THz. The measurement was recorded with a commercial ECOPS system.

A separate examination of the phase is left out here, because the sensitivity criterion in equation 1 in the main paper only uses the amplitude and not the phase. This approach also has the advantage that the setup for a measurement additionally evaluating the phase requires more precise adjustment, especially when operating a THz-TDS setup in reflection geometry, making an amplitude-based criterion easier to be applied. 
Based on the above agreement between measurement and simulation, the spectra used throughout this work, can be considered quite realistic.

\bibliography{Bibliography}

\end{document}